\documentstyle[12pt]{article}
\begin{document}

\setcounter{page}{1}

\begin{center}

{\LARGE \bf Calculations of Higher Twist Distribution Functions in the
MIT Bag Model} \\
\vspace{0.6cm}
A.~I.~ Signal \\
{\it Department of Physics, Massey University \\
Palmerston North, New Zealand \\
and \\
Institute for Theoretical Physics, University of Adelaide 
SA 5005, Australia}\\
\end{center}

\begin{abstract}

We calculate all twist-two, three and four parton distribution functions
involving two quark correlations using the wavefunction of the MIT bag
model. The distributions are evolved up to experimental scales and 
combined to give the various nucleon structure functions. Comparisons with 
recent experimental data on higher twist structure functions at moderate 
values of $Q^{2}$ give good agreement with the calculated structure functions.

\end{abstract}

\vspace{4cm}


\vspace{1cm}

Corresponding Author: 
Dr A I Signal
Department of Physics
Massey University, Private Bag 11-222
Palmerston North, New Zealand.

A.I.Signal@massey.ac.nz

Telephone: +64 6 350 4056
Fax: +64 6 354 0207

\newpage
\baselineskip=1.0cm
\section{Introduction}

Deep inelastic lepton - nucleon scattering (DIS) has been an important
tool in particle physics for more than twenty-five years. Recent high 
precision measurements by groups such as the NMC, SMC and various SLAC 
groups~\cite{NMC, SMC, E140, E142} are testing the limits of our theoretical 
knowledge of the structure of the nucleon. The SMC and the E143 groups 
\cite{SMC2, E143} have recently reported the first measurements of the 
transverse spin structure function $G_{2}(x, Q^{2})$
\footnote{Throughout this work I use capital letters {\it F, G} to denote 
the structure functions measured in experiment and lower case letters 
{\it f, g, h} to denote the various parton distributions.}
, which has a twist-three component $g_{T}$. In the near future we can also 
expect the HERMES experiment to investigate the spin dependent structure 
functions in some detail. Also the development of polarized beams at RHIC 
will lead to measurements of the chiral-odd distributions $h_{1,L}$, which 
are twist-two and twist-three respectively~\cite{JJ}.

Higher twist structure functions will also be important at the energies of
CEBAF and the proposed ELFE accelerator because of the phenomenon of
hadron-parton duality~\cite{BG,JU}, or `precocious scaling'~\cite{DGP}.
This predicts that at even moderate values of $Q^{2}$, higher twist
contributions to observables, such as the nucleon form factor, will be
non-negligible. Thus it is important to have some theoretical understanding
on the expected sizes and shapes of the higher twist structure functions.

In a recent paper, Ji~\cite{Ji} defined the 14 possible nucleon structure
functions in the standard model, and calculated each structure function in
terms of all the possible parton distribution functions up to the level of
twist-four (ignoring loop effects from QCD radiative corrections). The
parton distribution (or correlation) functions are defined in terms of
Fourier transforms of matrix elements of non-local quark and gauge fields
separated along the light-cone~\cite{JJ, Ji, EFP}. At present these matrix
elements cannot be calculated in QCD, however calculations of twist-two
matrix elements using the wavefunction of the MIT bag model have been
performed~\cite{SST, AS}, and these describe current data relatively well.
In this paper these calculations are extended to all the parton distribution
functions of twist-three and four containing only two quark fields.

\section{Parton Distribution Functions}

Following Ji~\cite{Ji}, let us consider a nucleon of mass $M$ with
momentum $P^{\mu}$ and polarisation vector $S^{\mu}$. If we choose the
nucleon to be moving along the $z$-axis with momentum $P$ we have

\begin{equation}
P^{\mu} = (\sqrt{M^{2} + P^{2}}, 0, 0, P).
\end{equation}
We introduce two orthogonal, light-like vectors $p$ and $n$

\begin{eqnarray}
p^{\mu} & = & \frac{1}{2}(\sqrt{M^{2} + P^{2}} + P)(1, 0, 0, 1), \nonumber \\
n^{\mu} & = & \frac{1}{M^{2}}(\sqrt{M^{2} + P^{2}} - P)(1, 0, 0, -1),
\end{eqnarray}
satisfying $P^{\mu} = p^{\mu} + M^{2}n^{\mu}/2$. We also decompose the
polarisation vector, $S = S_{\parallel} + M S_{\perp}$, where
\begin{equation}
S_{\parallel}^{\mu} = p^{\mu} - \frac{M^{2}}{2}n^{\mu},\;\;
S_{\perp}^{\mu} = (0, 1, 0, 0).
\end{equation}

A parton distribution with $k$ light-cone momentum fractions
$M(x_{1}, \ldots, x_{k})$ is defined via the matrix element
\begin{equation}
\int \prod_{i=1}^{k} \frac{d\lambda_{i}}{2\pi} \exp(i\lambda_{i}x_{i})
\langle PS | \hat{Q}(\lambda_{1}n, \ldots, \lambda_{k}n) | PS \rangle
= M(x_{1}, \ldots, x_{k}) \hat{T}(p, n, S_{\perp})
\end{equation}
where $\hat{Q}$ is a product of $k$ quark and gluon fields, and $\hat{T}$
is a Lorentz tensor.

If the mass dimension of the parton distribution $M(x_{i})$ is $d_{M}$ and 
that of $\hat{T}$ is $d_{T}$, then at large $Q^{2}$ the matrix element will 
behave as $\Lambda^{d_{M}}Q^{d_{T}}$, where $\Lambda$ is a soft QCD mass 
scale related to non-perturbative physics. A distribution with mass 
dimension $d_{M}$ is called a twist-$(d_{M}+2)$ distribution. As the
dimension of a physical observable is fixed, the higher $d_{M}$ becomes,
the higher the inverse power of hard momenta must become in the observable.
Thus a twist-$n$ parton distribution can only contribute to physical
observables of twist-$n$ or higher, which behave as $Q^{2-(n+a)}$, where
$a \geq 0$, in the $Q^{2} \rightarrow \infty$ limit. In the following we 
will often rescale a parton distribution by factors of $(M/\Lambda)$ to 
make it dimensionless, but the twist willl be unchanged. 

For scattering processes involving two quark fields we can define a quark
density matrix
\begin{equation}
M_{\alpha\beta}(x) = \int \frac{d\lambda}{2\pi} e^{i\lambda x}
\langle PS | \bar{\psi}_{\beta}(0) \psi_{\alpha}(\lambda n) | PS \rangle.
\end{equation}
From this it is possible to systematically generate the possible
distributions at a given twist. The twist-2 part of the density matrix can
be written
\begin{equation}
M(x)|_{\mbox{twist-2}} = \frac{1}{2} \not \! p f_{1}(x) +
\frac{1}{2}\gamma_{5} \not \!p (S_{\parallel} \cdot n) g_{1}(x) +
\frac{1}{2}\gamma_{5} \not \! S_{\perp} \not \! p h_{1}(x)
\end{equation}
where the three quark distribution functions $f_{1}$, $g_{1}$ and $h_{1}$
are defined
\begin{eqnarray}
f_{1}(x) & = & \frac{1}{2} \int \frac{d\lambda}{2\pi} e^{i\lambda x}
\langle P | \bar{\psi}(0) \not \! n \psi(\lambda n) | P \rangle, \nonumber \\
g_{1}(x) & = & \frac{1}{2}\int \frac{d\lambda}{2\pi} e^{i\lambda x}
\langle PS_{\parallel} | \bar{\psi}(0) \not \! n \gamma_{5} \psi(\lambda n)
| PS_{\parallel} \rangle, \nonumber \\
h_{1}(x) & = & \frac{1}{2}\int \frac{d\lambda}{2\pi} e^{i\lambda x}
\langle PS_{\perp} | \bar{\psi}(0) \not \! n \gamma_{5} \not \! S_{\perp}
\psi(\lambda n) | PS_{\perp} \rangle. \label{eq:tw2}
\end{eqnarray}
These represent the unpolarized quark density, the quark helicity density
and the quark transversity density~\cite{JJ} respectively. The light-cone
gauge, $A^{+} = 0$, has been chosen, so that the distributions are manifestly
gauge invariant.

Similarly at the twist-3 level, the appropriate portion of the quark
density matrix can be written
\begin{equation}
M(x)|_{\mbox{twist-3}} = \frac{\Lambda}{2} [ e(x) +
\frac{1}{2}(S_{\parallel} \cdot n) (\not \! p \not \! n - \not \! n \not \! p)
\gamma_{5} h_{L}(x) + \gamma_{5} \not \! S_{\perp} g_{T}(x) ]
\end{equation}
where $\Lambda$ is a soft mass scale in QCD. The three twist-3 distribution
functions are given by
\begin{eqnarray}
e(x) & = & \frac{1}{2\Lambda} \int \frac{d\lambda}{2\pi} e^{i\lambda x}
\langle P | \bar{\psi}(0) \psi(\lambda n) | P \rangle, \nonumber \\
h_{L}(x) & = & \frac{1}{2\Lambda}\int \frac{d\lambda}{2\pi} e^{i\lambda x}
\langle PS_{\parallel} | \bar{\psi}(0) \frac{1}{2} (\not \! p \not \! n -
\not \! n \not \! p)\gamma_{5} \psi(\lambda n) | PS_{\parallel} \rangle,
\nonumber \\
g_{T}(x) & = & \frac{1}{2\Lambda}\int \frac{d\lambda}{2\pi} e^{i\lambda x}
\langle PS_{\perp} | \bar{\psi}(0) \gamma_{5} \not \! S_{\perp} \psi(\lambda n)
| PS_{\perp} \rangle. \label{eq:tw3}
\end{eqnarray}

At twist-4 we have
\begin{equation}
M(x)|_{\mbox{twist-4}} = \frac{\Lambda^{2}}{4} [ \not \! n f_{4}(x) +
\not \! n \gamma_{5} (S_{\parallel} \cdot n) g_{3}(x) +
\not \! n \gamma_{5} \not \! S_{\perp} h_{3}(x) ]
\end{equation}
with the twist-4 distributions
\begin{eqnarray}
f_{4}(x) & = & \frac{1}{2 \Lambda^{2}} \int \frac{d\lambda}{2\pi}
e^{i\lambda x} \langle P | \bar{\psi}(0) \not \! p \psi(\lambda n)
| P \rangle, \nonumber \\
g_{3}(x) & = & \frac{1}{2 \Lambda^{2}}\int \frac{d\lambda}{2\pi}
e^{i\lambda x} \langle PS_{\parallel} | \bar{\psi}(0) \gamma_{5} \not \! p
\psi(\lambda n) | PS_{\parallel} \rangle, \nonumber \\
h_{3}(x) & = & \frac{1}{2 \Lambda^{2}}\int \frac{d\lambda}{2\pi}
e^{i\lambda x} \langle PS_{\perp} | \bar{\psi}(0) \gamma_{5} \not \! S_{\perp}
\not \! p \psi(\lambda n) | PS_{\perp} \rangle. \label{eq:tw4}
\end{eqnarray}

The quark field $\psi$ can be decomposed into `good' and `bad' components,
$\psi_{+}$ and $\psi_{-}$ respectively
\begin{equation}
\psi_{\pm} = P^{\pm} \psi, P^{\pm} + \frac{1}{2}\gamma^{\mp}\gamma^{\pm},
\gamma^{\pm} = \frac{1}{\sqrt{2}}(\gamma^{0} \pm \gamma^{3}).
\end{equation}
By inspection it can be seen that the twist-2 quark distributions involve
only the `good' component $\psi_{+}$, whereas the twist-3 distributions
involve mixing one `good' and one `bad' component, and the twist-4
distributions involve only the `bad' components.

The QCD equations of motion~\cite{KS}
\begin{eqnarray}
i\frac{d}{d\lambda} \psi_{-}(\lambda n) & = &
\frac{1}{2} \not \! n (-i\! \not \! D_{\perp} + m_{q}) \psi_{+}(\lambda n) , \\
-i\frac{d}{d\lambda} \bar{\psi}_{-}(\lambda n) & = &
\frac{1}{2} \bar{\psi}_{+}(\lambda n) (-i\! \not \! D_{\perp} + m_{q})
\not \! n ,
\end{eqnarray}
make it possible to eliminate the `bad' components from the twist-three and
four distributions, at the cost of introducing gluon fields into the matrix
elements. Because the model wavefunctions do not include gluon fields, we 
will not do this here. Also note that at twist-three and twist-four there
exist distributions involving the gluon field with two or three light-cone
momentum fractions, such as~\cite{Ji, EFP} 
\begin{equation}
E(x, y) = -\frac{1}{4\Lambda} \int \frac{d\lambda}{2\pi}
\frac{d\mu}{2\pi} e^{i\lambda x} e^{i\mu (y-x)} \langle P | \bar{\psi}(0)
i\! \not \! D_{\perp}(\mu n) \not \! n \psi(\lambda n) | P \rangle, \\
\end{equation}
and
\begin{equation}
B_{1}(x, y, z) = \frac{1}{2\Lambda^{2}} \int \frac{d\lambda}{2\pi}
\frac{d\mu}{2\pi} \frac{d\nu}{2\pi} e^{i\lambda x} e^{i\mu (y-x)}
e^{i\nu (z-y)}\langle P | \bar{\psi}(0) \not \! n i\! \not \!
D_{\perp}(\nu n) i\! \not \! D_{\perp}(\mu n) \psi(\lambda n) | P \rangle, \\
\end{equation}
which can be related to $e(x)$ and $f_{4}(x)$ by integrating over $y$ or
$y$ and $z$ respectively. However, as the MIT bag wavefunction has no
explicit gluon field, these distributions will be zero in the model. Also
the distributions with only one light-cone momentum fraction are of the
most interest for DIS and Drell-Yan processes.

Finally there exist distributions at the twist-four level involving four
quark operators eg
\begin{equation}
U^{s}_{1}(x, y, z) = \frac{1}{4\Lambda^{2}} \int \frac{d\lambda}{2\pi}
\frac{d\mu}{2\pi} \frac{d\nu}{2\pi} e^{i\lambda x} e^{i\mu (y-x)}
e^{i\nu (z-y)}\langle P | \bar{\psi}(0) \not \! n \psi(\nu n) \bar{\psi}(\mu n)
\psi(\lambda n) | P \rangle,
\end{equation}
which is a four quark light-cone correlation function. Calculating this
distribution by the method below would require the evaluation of the
overlap integral between the four quark fields over the bag volume, which
is expected to much smaller than the two quark overlap integral required
for the two quark correlation functions. Hence these will not considered 
further in this work. 

\section{Calculation of Quark Distributions}

At present no QCD wavefunction for the nucleon can be calculated. So in
order to make useful calculations of the quark distributions at any twist
it is necessary to the use the wavefunction from some phenomenological model
of the nucleon. The MIT bag model~\cite{MIT} is used here as it incorporates 
relativistic, light quarks, and also models confinement. It also has the 
further advantage that the wavefunction is simple and analytic. Other models 
could also be chosen~\cite{BS}.

The major problem in calculating the relevant matrix elements for the quark
distributions is ensuring that momentum conservation is obeyed throughout
the calculation, hence ensuring that the calculated distributions have the
correct support, i.e. they vanish for light-cone momentum fraction $x$
outside the interval $[0, 1]$ \cite{ST}. To guarantee momentum conservation, 
a complete set of intermediate states, $\sum_{m} |m \rangle \langle m|$, can 
be inserted into the matrix elements of the quark distributions
(eqns.~(\ref{eq:tw2}, \ref{eq:tw3}, \ref{eq:tw4})). Using translational
invariance of the matrix element, all the $n$ dependence can be moved 
into the argument of the exponential function. Then integrating over 
$\lambda$ gives a momentum conserving delta function. The twist-two quark
distributions then become
\begin{eqnarray}
f_{1}(x) & = & \frac{1}{\sqrt{2}} \sum_{m} \delta (p^{+}(1-x) - p_{m}^{+})
|\langle m | \psi_{+}(0) | P \rangle |^{2}, \nonumber \\
g_{1}(x) & = & \frac{1}{\sqrt{2}} \sum_{m} \delta (p^{+}(1-x) - p_{m}^{+})
[|\langle m | \hat{R} \psi_{+}(0) | PS_{\parallel} \rangle |^{2} -
|\langle m | \hat{L} \psi_{+}(0) | PS_{\parallel} \rangle |^{2}], \nonumber \\
h_{1}(x) & = & \frac{1}{\sqrt{2}} \sum_{m} \delta (p^{+}(1-x) - p_{m}^{+})
[|\langle m | \hat{Q}_{+} \psi_{+}(0) | PS_{\perp} \rangle |^{2} - \nonumber \\
&&\hspace{6cm}|\langle m | \hat{Q}_{-} \psi_{+}(0) | PS_{\perp} \rangle |^{2}],
\label{eq:t2}
\end{eqnarray}
where $\hat{R}(\hat{L})$ is the projection operator for right-(left-) handed
quarks $\hat{R}(\hat{L}) = (1 \pm \gamma_{5})/2$, and
$\hat{Q}_{\pm}$ is the projection operator
$\hat{Q}_{\pm} = (1 \pm \gamma_{5} \not \! S_{\perp})/2$, which
projects out eigenstates of the transversely projected Pauli-Lubanski
operator $\not \! S_{\perp}\gamma_{5}$ in a transversely projected nucleon.

The twist-four distributions are similar to those of twist-two, except they
involve `bad' components of the quark wavefunction
\begin{eqnarray}
f_{4}(x) & = & \frac{1}{\sqrt{2}} \sum_{m} \delta (p^{+}(1-x) - p_{m}^{+})
|\langle m | \psi_{-}(0) | P \rangle |^{2}, \nonumber \\
g_{3}(x) & = & \frac{1}{\sqrt{2}} \sum_{m} \delta (p^{+}(1-x) - p_{m}^{+})
[|\langle m | \hat{L} \psi_{-}(0) | PS_{\parallel} \rangle |^{2} -
|\langle m | \hat{R} \psi_{-}(0) | PS_{\parallel} \rangle |^{2}], \nonumber \\
h_{3}(x) & = & \frac{1}{\sqrt{2}} \sum_{m} \delta (p^{+}(1-x) - p_{m}^{+})
[|\langle m | \hat{Q_{+}} \psi_{-}(0) | PS_{\perp} \rangle |^{2} - \nonumber \\
&&\hspace{6cm}|\langle m | \hat{Q_{-}} \psi_{-}(0) | PS_{\perp} \rangle |^{2}].
\label{eq:t4}
\end{eqnarray}
Here the distributions have been rescaled to make them dimensionless, and
factors of $(M / \Lambda)^{2}$ have been absorbed in the twist-four part of
the density matrix, $M(x)|_{\mbox{twist-4}}$.

Note that the twist-two and twist-four distributions have a natural
interpretation in the parton model, where they are related to the
probability of finding a parton carrying fraction $x$ of the plus component
of momentum of the nucleon, and in the appropriate helicity or transversity
eigenstates. In the twist-three case, the distributions do not have a
similar interpretation in the parton model. However we can still guarantee
momentum conservation by introducing a complete set of intermediate states, 
$\sum_{m} |m \rangle \langle m|$, and then write the distributions in terms
of the matrix elements between nucleon states and intermediate states
\begin{eqnarray}
e(x) & = & \sum_{m} \delta (p^{+}(1-x) - p_{m}^{+})
\langle P | \psi^{\dagger}(0) | m \rangle_{a} (\gamma^{0})_{ab}
\langle m | \psi(0) | P \rangle_{b} , \nonumber \\
h_{L}(x) & = &  \sum_{m} \delta (p^{+}(1-x) - p_{m}^{+})
\langle PS_{\parallel} | \psi^{\dagger}(0) | m \rangle_{a} (\gamma^{3} 
\gamma^{5})_{ab} \langle m | \psi(0) | PS_{\parallel} \rangle_{b}, \nonumber \\
g_{T}(x) & = & \sum_{m} \delta (p^{+}(1-x) - p_{m}^{+})
\langle PS_{\perp} | \psi^{\dagger}(0) | m \rangle_{a} (\gamma^{0} \gamma^{5}
\not \! S_{\perp})_{ab}  \langle m | \psi(0) | PS_{\perp} \rangle_{b}
\label{eq:t3}
\end{eqnarray}
where the matrix indices have been shown explicitly. Again the distributions 
have been rescaled so that they are dimensionless, absorbing factors of 
$M / \Lambda$ into the density matrix $M(x)|_{\mbox{twist-3}}$.

The next step in the model calculation of the distributions is to form the
momentum eigenstates $|P\rangle$ and $|m\rangle$ from the static states of
the model. This can be done using either the Peierls-Yoccoz~\cite{PY}
projection, which gives a momentum dependent normalisation, or the
Peierls-Thouless~\cite{PT} projection, which leads to a more difficult
calculation, but which preserves Galilean invariance of the matrix
elements. Using either method, the correct support for the distributions is 
guaranteed by the above formalism. The distributions can then be calculated 
in terms of the Hill-Wheeler overlap integrals between the quark
wavefunctions~\cite{SST,AS}.

Using the wavefunction of a model also introduces a scale $\mu$ into the
calculated distribution functions. This is the scale at which the model
wavefunction is considered a good approximation to the true QCD
wavefunction, which is presently unknown. The natural scale for the bag
model, and most other phenomenological models employing light relativistic
quarks, is the typical transverse momentum of the quarks, 
$k_{T} \approx 400 \mbox{MeV}$. In order to
compare a calculated distribution function with experiment, the
calculated distribution needs to be evolved from the model scale up to the
experimental scale $Q^{2}$. This has previously been done using leading order
QCD evolution for the twist-2 distributions $f_{1}(x)$ and
$g_{1}(x)$~\cite{SST, AS}, with good agreement being obtained for a value
of $\mu$ in the region of 250 -- 500 MeV. This could be criticised on the
grounds that the strong coupling constant is not small in this region,
however calculations using next to leading order evolution\cite{StT} also
give good agreement with experiment for values of $\mu \approx 350 \mbox{MeV}$ 
and $\alpha_{S}(Q^{2} = \mu^{2}) \approx 0.75$.

For the higher twist distributions there do not yet exist comprehensive
calculations of the relevant anomalous dimensions for evolution of the
distribution functions. For the twist-three distributions $g_{T}$ and 
$h_{L}$ full calculations of the anomalous dimensions have appeared in 
the literature \cite{SV,KYU,KT}. For the lowest moments of these 
distributions ($n=3,4$), the leading order evolution gives similar results 
to naive power counting for $\sqrt{Q^{2}}$ in the region of 1 GeV. Thus we 
shall neglect QCD corrections to the evolution of the higher twist 
distributions, and use naive power counting to evolve these distributions.

In Figure 1 the nine twist-2, 3 and 4 distributions involving two quark 
correlation functions, calculated at the bag scale $\mu$, for a bag radius of 
0.8 fm are shown. The Peierls-Yoccoz projection has been used to form the 
momentum eigenstates $|p\rangle$ and $|m\rangle$, and only intermediate 
states containing two quarks have been considered. As is expected from
equations~ (\ref{eq:t2}, \ref{eq:t4}, \ref{eq:t3}), each of these
distribution functions are similar in magnitude to one another. It is worth
noting that the parton distributions satisfy the equalities
\begin{eqnarray}
f_{1}(x) + g_{1}(x) & = & 2h_{1}(x) \nonumber \\
e(x) + h_{L}(x) & = & 2g_{T}(x) \nonumber \\
f_{4}(x) + g_{3}(x) & = & 2h_{3}(x)
\end{eqnarray}
which are the lower bounds of Soffer's inequalities\cite{So}.

The evolution of the twist-3 distributions is particularly interesting
as the distributions have contributions from local operators of both
twist-3 and twist-2. Hence the moments of the twist-3 distributions 
are related to the moments of the twist-2 distributions
\cite{JJ,KYU,KT}
\begin{eqnarray}
{\cal M}_{n}[e] & = & \frac{m}{M} {\cal M}_{n}[f_{1}] + {\cal M}_{n}[e^{(3)}] 
\nonumber \\
{\cal M}_{n}[h_{L}] & = & \frac{2}{n+2}{\cal M}_{n}[h_{1}] + \frac{n}{n+2} 
\frac{m}{M}{\cal M}_{n-1}[g_{1}] + {\cal M}_{n}[h_{L}^{(3)}] \nonumber \\
{\cal M}_{n}[g_{T}] & = & \frac{1}{n+1} {\cal M}_{n}[g_{1}] - \frac{1}{2} 
\frac{m}{M}{\cal M}_{n-1}[h_{1}] + {\cal M}_{n}[g_{T}^{(3)}]
\end{eqnarray}
where $m$ is the quark mass, 
${\cal M}_{n}[f] = \int_{0}^{1} x^{n}f(x)dx$ is the $n$th moment of the 
distribution $f(x)$ and $f^{(3)}$ denotes the genuine twist-3 part of the
distribution $f(x)$. Note that when contributions from operators proportional 
to the quark mass or from twist-3 operators are ignored, the third of these 
relations is equivalent to the Wandzura-Wilczek relation \cite{WW}
\begin{equation}
g_{2}(x) = -g_{1}(x) + \int_{x}^{1}\frac{g_{1}(y)}{y}dy.
\end{equation}
In what follows we shall ignore terms directly proportional to the quark 
mass, as the masses of the $u$ and $d$ quarks are zero (or very small) 
in the MIT bag model. (If this discussion were extended to include strange 
quarks and anti-quarks then it would be important to keep terms in the
quark mass.)

We can invert the moments above to find the genuine twist-3 parts of
each distribution in terms of the calculated distributions and the 
twist-2 distributions
\begin{eqnarray}
e^{(3)}(x) & = & e(x) \nonumber \\
h^{(3)}_{L}(x) & = & h_{L}(x) - 2x \int_{x}^{1} \frac{dy}{y^{2}} h_{1}(y)
\nonumber \\
g^{(3)}_{T}(x) & = & g_{T}(x) - \int_{x}^{1} \frac{dy}{y} g_{1}(y). 
\label{eq:t3-2}
\end{eqnarray} 
This separation makes it possible to evolve the calculated distributions 
$e, h_{L}, g_{T}$ from the model scale up to experimental scales. The 
procedure I use for the evolution is to separate the calculated distributions 
at the model scale $Q^{2}=\mu^{2}$ into twist-2 and twist-3 parts using 
eqn.~(\ref{eq:t3-2}). The twist-2 parts are then evolved using the 
Gribov-Lipatov-Altarelli-Parisi (GLAP) equation \cite{GLAP} to leading order, 
whereas the twist-3 parts are evolved according to naive power counting 
$f^{(3)}(Q^{2}) \sim 1/\sqrt{Q^{2}}$. The model scale has been chosen as 
$\mu = 0.4$ GeV, and in Figure 2 the full twist-3 distributions at 
$Q^{2}$ = 1 and 10 GeV$^{2}$ are shown. Comparing between $e(x, Q^{2})$, 
which has no twist-2 part, and $h_{L}(x, Q^{2})$ and $g_{T}(x,Q^{2})$ enables 
us to see how the twist-2 parts of the latter two distributions dominate at 
higher $Q^{2}$.

In fact it is probably not a particularly good approximation to evolve 
$e(x, Q^{2})$ according to naive power counting. While a calculation of the
anomalous dimensions for the operators contributing to $e(x)$ has yet to 
be done, we can surmise a few facts about the lowest moments, including that 
the lowest moment of $e(x)$ will have a small anomalous dimension despite the 
corresponding operator being twist-3. The relevant local operators for 
$e(x)$ in the operator product expansion are 
\begin{equation}
O^{\mu_{1} \mu_{2} \ldots \mu_{n}} = {\cal S}_{n} \bar{\psi} (iD^{\mu_{1}})
(iD^{\mu_{2}}) \ldots (iD^{\mu_{n}}) \psi - \mbox{traces}
\end{equation}
where ${\cal S}_{n}$ symmetrizes over the Lorentz indices 
$\mu_{1}, \ldots \mu_{n}$. The relevant matrix elements of these operators 
are
\begin{equation}
\langle P | O^{\mu_{1} \mu_{2} \ldots \mu_{n}} | P \rangle = 
2 M e_{n} P^{\mu_{1}} P^{\mu_{2}} \ldots P^{\mu_{n}} - \mbox{traces}.
\end{equation}
Standard techniques give the sum rules
\begin{equation}
e_{n} = \int dx x^{n} e(x).
\end{equation}
In particular, for $n = 0$ we have
\begin{equation}
\int dx e(x) = \frac{1}{2M} \langle P | \bar{\psi}\psi | P \rangle
\label{eq:e0}
\end{equation}
which is related to the nucleon $\sigma$ term, 
$\langle P |\frac{1}{2}(m_{u}+m_{d})(\bar{u}u + \bar{d}d) |P\rangle$. The
$\sigma$ term is renormalization point invariant in QCD, which implies that
the matrix element in eqn.~(\ref{eq:e0}) will have an anomalous dimension of 
the same magnitude as that of the quark mass operator. Thus the zeroth 
moment of $e(x)$ will change quite slowly with $Q^{2}$, whereas we expect 
that the higher moments should scale approximately as $1/\sqrt{Q^{2}}$. 
Thus the distribution $e(x,Q^{2})$ will tend to increase at low $x$ and 
decrease at high $x$ as $Q^{2}$ increases
\footnote{On completion of this work I learned of a calculation of the 
anomalous dimensions of $e(x,Q^{2})$ \cite{KN} which confirms these 
speculative remarks, in particular the anomalous dimensions of the operators 
corresponding to the second and third moments of $e$ are similar in magnitude 
to the anomalous dimensions of twist-2 distributions.}.

At twist-4 level there is no mixing of twist-2 and twist-3 operators with 
the local twist-4 operators \cite{JS}, which simplifies the analysis 
somewhat. Of course there are a large number of local operators of 
twist-4 for each distribution function, and in the vast majority of cases 
their anomalous dimensions have not been calculated. Thus to evolve the 
twist-4 distributions $f^{4}(x)$ naive power counting, 
$f^{4}(x, Q^{2}) \sim 1/Q^{2}$ is again used. In Figure 3 we see the twist-4 
distributions at $Q^{2} =$ 1 and 10 GeV$^{2}$, where again the bag scale has 
been taken as $\mu = 0.4$ GeV.

Finally, the twist-two transversity distribution $h_{1}(x)$ will be of 
interest at scales above 1 GeV$^{2}$. The anomalous dimensions of the 
twist-2 operators have been calculated \cite{AM}, so the evolution of the 
parton distribution from the bag scale is fairly straightforward. The results 
of this evolution up to $Q^{2} = 1$ GeV$^{2}$ and $Q^{2} = 1o$ GeV$^{2}$ are 
shown in Figure 4.

\section{Nucleon Structure Functions}

The parton distribution functions calculated above can be measured in 
various processes, most notably deep inelastic lepton-nucleon scattering 
(DIS) and Drell-Yan (DY) processes. In this section we consider how the parton 
distributions combine to give the nucleon structure functions which can be 
measured in DIS, and those structure functions involving higher twist 
distributions which have been measured are compared with the calculated 
structure functions.

In the Bjorken limit ($Q^{2}, \nu \rightarrow \infty$, 
$x= Q^{2}/2(P.q)$ fixed)
the nucleon tensor $W^{\mu\nu}$ can be expressed in terms of 14 structure 
functions \cite{Ji}
\begin{eqnarray}
W^{\mu\nu} & = & \left(-g^{\mu\nu} + \frac{q^{\mu}q^{\nu}}{q^{2}} \right)
F_{1}(x) + \hat{p}^{\mu}\hat{p}^{\nu} \frac{F_{2}(x)}{\nu} \nonumber \\ 
&& + q^{\mu}q^{\nu} \frac{F_{4}(x)}{\nu} + (p^{\mu}q^{\nu}+p^{\nu}q^{\mu}) 
\frac{F_{5}(x)}{2\nu} \nonumber \\ 
&& -i\epsilon^{\mu\nu\alpha\beta}q_{\alpha}\left(p_{\beta} \frac{G_{1}(x)}{\nu}
+ MS_{\perp\beta} \frac{G_{T}(x)}{\nu} \right) \nonumber \\ 
&& +i\epsilon^{\mu\nu\alpha\beta} Mp_{\alpha}S_{\perp\beta} 
\frac{G_{3}(x)}{\nu} +i\epsilon^{\mu\nu\alpha\beta} q_{\alpha}p_{\beta} 
\frac{F_{3}(x)}{2\nu} \nonumber \\ 
&& + \left(-g^{\mu\nu} + \frac{q^{\mu}q^{\nu}}{q^{2}} \right)
a_{1}(x) + \hat{p}^{\mu}\hat{p}^{\nu} \frac{a_{2}(x)}{\nu} \nonumber \\ 
&& + q^{\mu}q^{\nu} \frac{a_{4}(x)}{\nu} + (p^{\mu}q^{\nu}+p^{\nu}q^{\mu}) 
\frac{a_{5}(x)}{2\nu} \nonumber \\ 
&& + M(S_{\perp}^{\mu}\hat{p}^{\nu} + S_{\perp}^{\nu}\hat{p}^{\mu}) 
\frac{b_{1}(x)}{2\nu} + M(S_{\perp}^{\mu}p^{\nu} + S_{\perp}^{\nu}p^{\mu}) 
\frac{b_{2}(x)}{2\nu}, 
\end{eqnarray} 
where the last seven structure functions ($F_{3}, a_{1,2,4,5}, b_{1,2}$) are 
related to parity-violating processes involving the weak interaction. I have 
introduced the shorthand notation 
$\hat{p}^{\mu}=p^{\mu}-\frac{p.q}{q^{2}}q^{\mu}$.

It is conventional to introduce longitudinal structure functions, describing 
the scattering when the exchanged vector boson is longitudinally polarised
\begin{eqnarray}
F_{L}(x) & = & F_{2}(x)\left(1 + \frac{4M^{2}x^{2}}{Q^{2}}\right) 
- 2x F_{1}(x) \nonumber \\
a_{L}(x) & = & a_{2}(x)\left(1 + \frac{4M^{2}x^{2}}{Q^{2}}\right) 
- 2x a_{1}(x).
\end{eqnarray}
According to the Callen-Gross relation \cite{CG} both $F_{L}$ and $a_{L}$ 
vanish in the Bjorken limit, and, ignoring QCD radiative corrections and 
nucleon mass effects, both are twist-4 structure functions. 

The fourteen structure functions can all be expressed in terms of parton 
distribution functions \cite{Ji, EFP, JS}. In the following all 
contributions from parton distributions with two or more light-cone 
fractions (ie distributions involving two quark fields plus one or two 
gluon fields and distributions involving four quark fields) will be dropped 
and the discussion is limited to the distributions calculated above. We also 
keep terms linear in the quark masses. The electroweak current $J^{\mu}(\xi)$ 
of the quarks coupling to vector bosons is taken to be 
\begin{equation}
J^{\mu}(\xi) = \bar{\psi}(\xi) \gamma^{\mu} (g_{v} + g_{a}\gamma_{5}) 
\psi(\xi)
\end{equation}
where the vector and pseudo-vector couplings, $g_{v}$ and $g_{a}$ take the 
values of the standard model. Also, because weak currents can change quark 
flavour, the mass of a quark in the initial state is denoted as $m_{i}$, and 
the mass of a quark in the final state as $m_{f}$.

For unpolarised scattering there are five measureable structure functions. 
$F_{1}$ and $F_{2}$ (or $F_{L}$) describe electromagnetic scattering, 
$F_{3}$ is related to parity violating weak scattering, and $F_{4}$ and 
$F_{5}$ are related to the scattering of non-conserved currents. We have:
\begin{eqnarray} 
F_{1}(x) & = & \frac{1}{2} \sum_{q} (|g_{vq}|^{2} + |g_{aq}|^{2}) 
f_{1}^{q}(x) \nonumber \\ 
&& - \frac{M}{Q^{2}} \sum_{q} m_{f}(|g_{vq}|^{2} - |g_{aq}|^{2}) x e^{q}(x) \\
F_{L}(x) & = & \frac{M^{2}}{Q^{2}} \sum_{q} (|g_{vq}|^{2} + |g_{aq}|^{2}) 
4 x^{3} f_{4}^{q}(x) \nonumber \\ 
&& - \frac{M}{Q^{2}} \sum_{q} \left[(m_{i}+m_{f})|g_{vq}|^{2} 
+ (m_{f}-m_{i})|g_{aq}|^{2}\right] 4 x^{2} e^{q}(x) \\
F_{3}(x) & = & \sum_{q} (-1)^{q} (|g_{vq}|^{2} + |g_{aq}|^{2}) f_{1}^{q}(x) 
\end{eqnarray}
where $(-1)^{q}$ is +1 for quarks and -1 for antiquarks. Finally
\begin{eqnarray}
F_{4}(x) & = & \frac{M}{Q^{2}} \sum_{q} \left[|g_{vq}|^{2}(m_{i}-m_{f}) + 
|g_{aq}|^{2}(m_{i}+m_{f})\right] e^{q}(x) \label{eq:F4} \\
F_{5}(x) & = & \frac{2Mx}{Q^{2}} \sum_{q} \left[|g_{vq}|^{2}(m_{i}-m_{f}) + 
|g_{aq}|^{2}(m_{i}+m_{f})\right] e^{q}(x). \label{eq:F5}
\end{eqnarray}
Thus $F_{L}$, $F_{4}$ and $F_{5}$ are twist-four structure functions. 

For scattering from a longitudinally polarised nucleon we have five structure 
functions:
\begin{eqnarray}
G_{1}(x) & = & \frac{1}{2} \sum_{q} (|g_{vq}|^{2} + |g_{aq}|^{2}) 
g_{1}^{q}(x) \\ 
a_{1}(x) & = & \frac{1}{2} \sum_{q} (-1)^{q} (|g_{vq}|^{2} + |g_{aq}|^{2}) 
g_{1}^{q}(x) \\ 
a_{L}(x) & = & \frac{2Mx}{Q^{2}} \sum_{q} (-1)^{q} 
(|g_{vq}|^{2} + |g_{aq}|^{2})\left[-2Mx^{2}g_{3}^{q}(x) + 
2m_{i}x h_{L}^{q}(x) \right] \\ 
a_{4}(x) & = & \frac{M}{Q^{2}} \sum_{q} (-1)^{q} (|g_{vq}|^{2} + |g_{aq}|^{2}) 
m_{i} x h_{L}^{q}(x) \\ 
a_{5}(x) & = & \frac{2Mx}{Q^{2}} \sum_{q} (-1)^{q} 
(|g_{vq}|^{2} + |g_{aq}|^{2}) m_{i} x h_{L}^{q}(x). 
\end{eqnarray}
The twist-2 distribution $a_{1}$ is the longitudinally polarised analogue of 
$F_{3}$, and if it were measured the role of the axial anomaly in the 
interpretation of measurements of $G_{1}$ would be much clearer\cite{BT}. 
The structure functions $a_{L,4,5}$ are all twist-4, with $a_{4,5}$ related 
to non-conserved currents. 

For scattering from a transversely polarised nucleon there are four 
structure functions:
\begin{eqnarray}
G_{T}(x) & = & \frac{1}{2} \sum_{q} (|g_{vq}|^{2} + |g_{aq}|^{2}) 
g_{T}^{q}(x) \\ 
G_{3}(x) & = & \frac{1}{2M} \sum_{q} \left[|g_{vq}|^{2}(m_{f}-m_{i}) - 
|g_{aq}|^{2}(m_{i}+m_{f}) \right] h_{1}^{q}(x) \label{eq:G3} \\ 
b_{1}(x) & = & \sum_{q} (-1)^{q} (|g_{vq}|^{2} + |g_{aq}|^{2}) 
x g_{T}^{q}(x) \\ 
b_{2}(x) & = & \frac{1}{M} \sum_{q} (-1)^{q} (|g_{vq}|^{2} + |g_{aq}|^{2}) 
m_{i} h_{1}^{q}(x).
\end{eqnarray} 
These four structure functions are twist-3. $G_{T} = G_{1}+G_{2}$ is related 
to the transverse spin of the quarks and antiquarks, $b_{1}$ is its 
parity-violating partner, while $G_{3}$ and $b_{2}$ are related to 
non-conserved currents.

Many experiments over the years have determined $F_{1}$, $F_{3}$ and 
$G_{1}$ to good precision, and in the near future $G_{T}$ should be 
measured to similar precision. $F_{L}$ has also been measured to a lower 
degree of precision, thus it should be possible for us to extract 
estimates of the parton distributions $g_{T}$ and $f_{4}$. Unfortunately 
there is little hope of determining any of the other parton distributions 
from DIS. From the expressions for $F_{4,5}$, eqs.~(\ref{eq:F4}, \ref{eq:F5}) 
one might hope that $e(x)$ would be measurable in heavy quark production.
However because $|g_{vq}|=|g_{aq}|$ the terms in $m_{f}$ cancel
\footnote{This cancellation does not occur when the exchange boson is a Z 
boson, so there may exist a possibility to measure $F_{4,5}$ at HERA}. 
A similar cancellation occurs in eqn.~(\ref{eq:G3}) making a determination of 
$h_{1}(x)$ very difficult. However the possibility exists to measure 
$h_{1,L}(x)$ and $e(x)$ in polarised Drell-Yan processes \cite{JJ} which 
may be possible at RHIC.

Previous calculations \cite{SST, StT} have shown good agreement between 
experimental data for $F_{1}(x)$ and $F_{3}(x)$ and the model predictions. 
We are now in a position to compare the model prediction for $F_{L}(x)$ with 
experimental data. Experimentally the ratio $R = F_{L}/2xF_{1}$ has been 
measured \cite{E140}, and there is good evidence for a twist-4 contribution 
\cite{SGM+, BRY} at medium and high $x$ and $Q^{2}<10$ GeV$^{2}$. Before we 
can compare with the data we should also take into account effects from 
perturbative QCD which give a non-zero $F_{L}$ at leading and next-to-leading 
order, and effects from the target nucleon mass. In the simplest version of 
the bag model the twist four contribution to $R$ for electromagnetic 
scattering is given by
\begin{equation}
R^{(4)}(x, Q^{2})  =  \frac{8 M^{2}x^{2}}{Q^{2}}
\frac{f_{4}(x,Q^{2})}{f_{1}(x,Q^{2})} 
\end{equation}
In Figure 5 we show our calculated $R^{(4)}(x, Q^{2})$ at $Q^{2} = 2$ and 
5 GeV$^{2}$. We also show the experimental values of $R$ and various 
perturbative QCD calculations, including target mass effects, and the effect 
of adding the twist-4 contribution from the bag model at these scales. Because 
$R^{(4)}(x, Q^{2})$ effectively scales as $f_{4}(x, \mu^{2})/Q^{4}$, we 
calculate a large correction to the perturbative calculations at only low values 
of $Q^{2}$, however this twist-4 correction is in the right direction, and 
for low $Q^{2}$ gives a good fit to the data at medium $x$ values. 

If the data at low $Q^{2}$ improves in accuracy, it may be possible to perform 
an independent check on our value of the bag scale, $\mu$, from a fit of our 
calculated parton distribution $f_{4}$ with the data at various values of 
$Q^{2}$. 

The other higher twist structure function that has been measured is 
$G_{2}(x) = G_{T}(x) - G_{1}(x)$ \cite{E143, SMC2}. In Figure 6 the 
experimental data for $G_{2}^{p}(x)$ is compared with the calculated structure 
function at $Q^{2} = 5$ GeV$^{2}$. The calculation is in reasonable agreement 
with the data. In particular the calculated $G_{2}(x)$ crosses the $x$ axis 
in the right region. As this crossing is determined by the interplay between 
the twist-3 part of $G_{2}(x)$, given by $g_{T}^{(3)}(x)$ and the twist-2 portion 
$-g_{1}(x) + \int_{x}^{1} dy g_{1}(y)/y$, this gives us some 
confidence in our understanding of the relations between the twist-2 and 
twist-3 distributions. It will be interesting to check our model calculations 
against experimental at other values of $Q^{2}$. As $Q^{2}$ increases we 
predict that the crossing point $x_{0}$ where $G_{2}(x_{0}) = 0$ will move 
to larger $x$. 

It is also interesting to note that while the bag model calculation for the 
structure function $G_{1}^{p}(x)$ is not particularly good, presumably 
because the calculation does not include effects arising from the axial 
anomaly\cite{SST}, in the case of $G_{2}^{p}(x)$ the agreement with 
the data is better, probably because of the cancellation between the two terms 
in the twist-2 part of $G_{2}^{p}(x)$.

Finally in Table 1 the calculated values for some of moments of the 
twist-2 and twist-3 parts of $G_{2}^{p}(x)$ are displayed and compared with 
values from other calculations and those extracted from the experimental data.

Before concluding, there are some limitations to these calculations which 
should be borne in mind. There is no flavour dependence of the calculated 
parton distributions, so the distributions and structure functions shown 
above should be taken as applying to scattering from an isoscalar target. 
For our comparison of $R$ this is not important, as experiment has shown no 
target dependence for $R$ \cite{E140}, but for $G_{2}$ the flavour dependence 
will be as important as for $G_{1}$\cite{SST, SHT}. In the model, the 
flavour dependence of the parton distributions comes from the $SU(6)$ 
spin-flavour part of the wavefunction, and the one gluon exchange mechanism 
means that the two-quark intermediate state must be either a scalar (S=0) 
or vector (S=1) with masses $m_{s}$ and $m_{v} > m_{s}$, respectively 
\cite{CT, SST}. Furthermore no account has been taken of contributions to the 
parton distributions from intermediate states consisting of four quarks. 
However these contributions have only a small effect on valence 
distributions, and this is all in the small $x$ region \cite{SST}, which is 
not of great importance here. 

An important correction to the parton distributions will come from mesonic 
contributions via the so-called Sullivan process \cite{Sul}, where meson 
($\pi, \rho, K \ldots$) exchange between the virtual photon and the 
nucleon occurs. These processes are known to lead to important modifications 
in the parton distributions at twist-2 \cite{SMST,SHT}, particularly for 
$g_{1}(x)$, and there is no reason to suppose that the higher twist 
distributions should be unaffected. This should be important in the model 
calculation of the structure function $G_{2}$.

\section{Summary}

In this paper we have discussed the various parton distribution functions up
to twist four in the context of the MIT bag model. The discussion has been 
limited to those distributions involving two-quark correlation functions, 
however these include effects from the bag boundary, which is a simple model 
of the effects of the gluon field. From the twist-3 distributions, which 
involve quark-gluon correlations, we have calculated the twist-3 structure 
function, $G_{2}(x)$, and found the calculation in good agreement with the 
experimental data, which gives us some confidence in using the bag as a model 
of the quark-gluon dynamics involved in DIS. 

Four-quark correlation functions are expected to be small as they involve the 
calculation of Hill-Wheeler-type overlap functions for four quarks in the bag. 
Also these distributions are not directly accessible to experiment. However 
in future work we shall look at the corrections four-quark distributions make 
to the two-quark distributions calculated here. 

At twist-four we have seen that the calculated distribution $f_{4}(x)$ gives 
a contribution to the experimentally measured $R = F_{L}/2xF_{2}$. When added 
to the contributions from perturbative QCD and target mass effects, the 
calculation gives good agreement with the data at $Q^{2} \sim 1$ GeV$^{2}$ 
an dmedium values of $x$. 

Some improvements can be made to our calculations. On the model side we can 
add flavour dependence to our distributions, and we can also add the important 
effects of the pion cloud. Also we can improve our evolution techniques for 
the distribution functions. For the twist-2 distributions next-to-leading-order 
QCD corrections are available. At twist-3 there have been recent developments 
in the calculations of anomalous dimensions of the relevant operators, so that 
a correct leading order QCD approach is possible, though this will be very 
difficult.

I would like to thank the Institute for Theoretical Physics, University of 
Adelaide, for hospitality and support while this work was in completion. I 
would also like to thank Tony Thomas, Fernando Steffens and Steve Shrimpton 
for valuable comments. This work was supported in part by the Australian 
Research Council.

\newpage

\parindent=0.0cm

\section*{Figure Captions}

{\bf Figure 1.} The parton distribution functions calculated at the bag 
scale $Q^{2} = \mu^{2}$. (a) The twist-two distributions 
$f_{1}(x), g_{1}(x)$ and $h_{1}(x)$. (b) The twist-three distributions 
$e(x), h_{L}(x)$ and $g_{T}(x)$. (c) The twist-four distributions 
$f_{4}(x), g_{3}(x)$ and $h_{3}(x)$.

{\bf Figure 2.} The twist-three parton distributions at 
(a) $Q^{2} = 1$ GeV$^{2}$ and (b) $Q^{2} = 10$ GeV$^{2}$. 

{\bf Figure 3.} The twist-four parton distributions at 
(a) $Q^{2} = 1$ GeV$^{2}$ and (b) $Q^{2} = 10$ GeV$^{2}$.

{\bf Figure 4.} The transversity dependent twist-two parton distribution
$h_{1}(x, Q^{2})$ at $Q^{2} = 1$ GeV$^{2}$ and $Q^{2} = 10$ GeV$^{2}$.
 
{\bf Figure 5(a).} The twist-four distribution $R^{(4)}(x, Q^{2})$ 
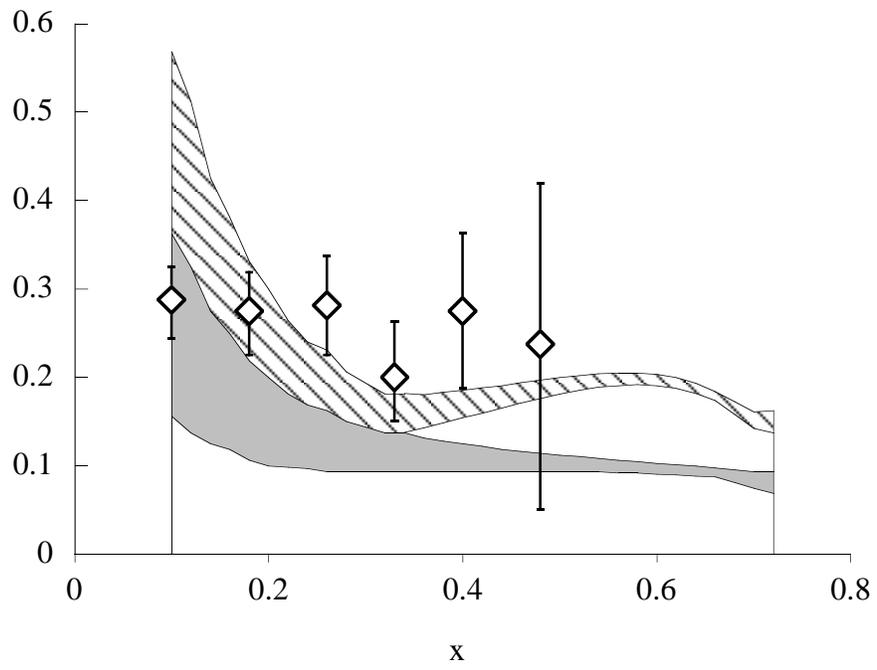
calculated at $Q^{2} = 2$ GeV$^{2}$ and $Q^{2} = 5$ GeV$^{2}$.

{\bf Figure 5(b).} Experimental data \cite{E140} on $R$ and QCD calculations 
with (upper band) and without (lower band) the twist-four contribution 
$R^{(4)}(x)$ at $Q^2 = 2$ GeV$^{2}$. 

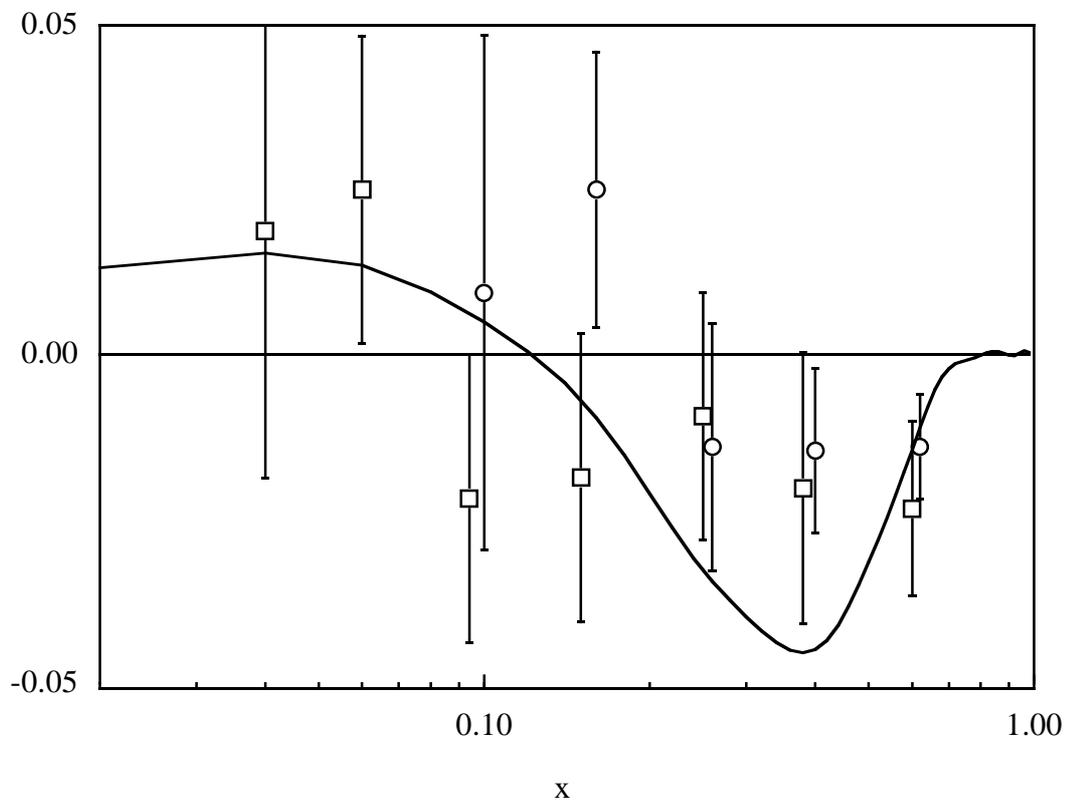
{\bf Figure 6.} Comparison of experimental data \cite{E143} on $G_{2}^{p}(x)$ 
with this calculation. 

\newpage
\begin{table}
\centering
\begin{tabular}{|l|ccc|} \hline \hline 
& ${\cal M}_{2}[G_{2}^{p}]$ & ${\cal M}_{4}[G_{2}^{p}]$ & 
${\cal M}_{6}[G_{2}^{p}]$ \\ \hline
This work & $-5.9\times 10^{-3}$ & $-1.2\times 10^{-3}$ & $-0.3\times 10^{-3}$ \\ 
C.M. Bag Model\cite{Song} & $-1.2\times 10^{-3}$ & & \\
Lattice QCD\cite{Gock} & $(-26.1\pm 3.8)\times 10^{-3}$ & & \\
Data\cite{E143} & $(-6.3\pm 1.8)\times 10^{-3}$ & $(-2.3\pm 0.6)\times 10^{-3}$ & 
$(-1.0\pm 0.3)\times 10^{-3}$ \\ \hline \hline
\end{tabular}
\vspace{2cm}

\begin{tabular}{|l|ccc|} \hline \hline
& $d_{2}^{p}$ & $d_{4}^{p}$ & $d_{6}^{p}$ \\ \hline
This work & $6.3\times 10^{-3}$ & $1.3\times 10^{-3}$ & $0.3\times 10^{-3}$ \\ 
C.M. Bag Model & $-17.4\times 10^{-3}$ & & \\
Lattice QCD & $(-48\pm 5)\times 10^{-3}$ & & \\
QCD Sum Rule\cite{Stein} & $(-6\pm 3)\times 10^{-3}$ & & \\
Data\cite{E143} & $(5.4\pm 5)\times 10^{-3}$ & $(0.7\pm 1.7)\times 10^{-3}$ & 
$(0.1\pm 0.8)\times 10^{-3}$ \\ \hline \hline
\end{tabular}
\end{table}

\vspace{2cm}
{\bf Table 1.} Comparison of various calculations with experimental data for the 
moments of $G_{2}^{p}$ and the twist-3 matrix elements 
$d_{n}^{p} = 2(\frac{n+1}{n}{\cal M}_{n}[G_{2}^{p}] + {\cal M}_{n}[G_{1}^{p}]) 
= 2 \frac{n+1}{n}{\cal M}_{n}[G_{T}^{(3)p}]$.

\end{document}